\documentclass[
prl,twocolumn,superscriptaddress,showpacs,groupedaddress,
 amsmath,amssymb,
 aps,
]{revtex4-1}

\usepackage{graphicx}
\usepackage{dcolumn}
\usepackage{bm}

\begin{document}

\preprint{APS/123-QED}

\title{Nucleation, solvation and boiling of helium excimer clusters}

\author{$^1$Luis~G.~Mendoza~Luna, $^1$Nagham~M.~Siltagh, $^1$Mark~J.~Watkins,
  $^2$Nelly Bonifaci, $^2$Fr{\'e}d{\'e}ric Aitken, $^1$Klaus von Haeften}
\email[]{kvh6@le.ac.uk} 
\affiliation{$^1$University of Leicester, Department of Physics \& Astronomy, Leicester, LE1 7RH, United Kingdom}
\affiliation{$^2$G2ELab-CNRS, Equipe MDE 25, Av. des Martyrs, BP 166, 38042
  Grenoble Cedex 9, France}





\date{\today}

\begin{abstract}
Helium excimers generated by a 
corona discharge were investigated in the gas and normal liquid phases
of helium as a function of 
temperature and pressure between 3.8 and 5.0~K and 0.2 and 5.6~bar. 
Intense fluorescence in the visible region showed the rotationally
resolved $d^3\Sigma_u^+ \rightarrow b^3\Pi_g$ transition of
He$_2^*$. 
With increasing pressure, the rotational lines merged into single
features. The observed pressure dependence of linewidths, shapes and lineshifts
established phases of coexistence and separation of excimer-helium mixtures,
providing detailed insight into nucleation, solvation and 
boiling of He$_2^*$-He$_n$ clusters. 
\end{abstract}

\pacs{Valid PACS appear here}
\maketitle

The ability to resolve rotational lines at low temperatures provides exceptional
sensitivity for investigating molecule-solvent interactions.
For example, infrared spectra of molecules embedded in 0.4~K cold
$^4$He droplets \cite{hartmann1995} 
display sharp, discrete lines, indicating both free rotation and the
superfluidity of the $^4$He droplets \cite{toennies2004}. A decreased
B-constant and a degree of line broadening reflect an interaction of the
molecule with the normal component of helium at 
the interface \cite{kwon1999,kwon2000} with the otherwise superfluid
helium. 
Mixed $^3$He-$^4$He droplets with more than 60 $^4$He atoms are
also superfluid and have an even lower temperature of only 0.15~K. Remarkably, in
this ultracold environment the linewidth of the rotational features is
three times \cite{toennies2004} {\em smaller} than in {\em pure} $^4$He
droplets \cite{grebenev1998,sartakov2012}.

In view of these exciting features and distinct temperature effects, this
would seem a very promising starting point from which to perform rotationally resolved spectroscopy
with control over a wide range of temperatures, beyond 0.15 and 0.4~K,
particularly where additional control over
pressure might also be possible. While this control is difficult to  
accomplish in helium droplets in free beams, it 
is quite readily possible using bulk helium. However, 
embedding single foreign molecules directly into bulk helium is
challenging because at the temperatures of liquid helium all other 
substances are themselves frozen, \cite{kiryukhin1997} and sophisticated techniques
are required to achieve this \cite{lebedev2010,khmelenko2013,popov2015}. 
Short-lived helium excimers (He$_2^*$) as single-molecule probes
represent an alternative means of investigation; they have been used both recently
\cite{benderskii1999,mckinsey2005,rellergert2008,guo2009,zmeev2013,guo2014} 
and in the past to probe the bulk phases of 
helium by imaging \cite{gao2015} and spectroscopy. Dennis et al.\ have bombarded 1.7~K
cold superfluid helium with electrons and observed fluorescence in the visible
spectral range, which originated from transitions between various electronically
excited singlet and triplet states of He$_2^*$
\cite{dennis1969}. Despite an environment of superfluid
helium, the spectra did not show
discrete rotational lines but rather features that resembled the rotational envelope of P,
Q and R transitions. However, in a similar experiment, Hill, Heybey and Walter
observed discrete rotational lines in the transient absorption spectrum
of He$_2^*$ \cite{hill1971} of 1.7~K cold superfluid helium. These lines
were shifted from their associated emission-in-helium and gas phase
values, though changes in the effective moment of inertia were not
reported. Li et al.\ excited normal-liquid helium with a corona discharge
and observed the fluorescent emission of He$_2^*$, similar to Dennis et al., 
but in this instance sharp, discrete lines were observed \cite{li2009}. In all
these experiments, however, the effects of temperature and the pressure have not
been addressed.

To resolve this apparent contradiction and to advance the understanding of
solvation in liquids
we have recorded fluorescence spectra of He$_2^*$ 
over a wide range of hydrostatic pressures between 0.2 and 5.6~bar, and
at temperatures between 3.8 and 5.0~K, covering the gas and normal liquid
phases of helium. A corona discharge was chosen for 
electronic excitation because it can operate over a very wide range of
pressures. 
At low pressures, spectra were observed showing discrete rotational
lines. Upon increasing the
pressure, the lines broadened and ultimately merged into broad features; 
the lines were also shifted in frequency. Analysis of the pressure dependence
of this line shift revealed that He$_2^*$ excimers exist in locally heated `gas
pockets' as well as in solvated states, both of which contribute to the 
associated spectrum. Simulations of the gas phase spectrum show that about 20\% of
the molecules are in a solvated state, regardless of pressure and temperature.
At 3.8~K, features of solvated He$_2^*$ appear at pressures slightly
lower than the saturated vapour pressure (SVP) of {\em pure} helium, indicating that
clusters of excimers and ground state helium atoms form at conditions
where helium is not yet in the liquid state. The points of cluster
formation, indicating the 
SVP curve of the mixed phase of He$_2$ and He, were found to cross the SVP
curve of {\em pure} helium at higher temperatures. Consequently, the
mixed He$_2$-He system boils in this region before {\em pure} helium,
giving rise to the formation of localised `gas pockets'.

The experimental setup was conceptually similar to that described by Li et al.\ 
\cite{li2009}, but employed a closed-cycle Oxford Instruments Heliox AC-V He3
cryostat. A micro-discharge cell, with an internal volume of 4~ml and made of
oxygen-free copper (OFHC), was attached to 
the $^3$He stage, but it was found that sufficient cooling power could
only be provided after bridging the second stage with copper strips. 
With these bridges, the cell could reach a minimum
temperature of 3.2~K, which was measured by a calibrated Cernox resistor
within $\pm$3~mK \footnote{On the SVP, the temperature of the helium inside the cell
  deviated by no more than 0.1~K on the sensor readout}. The cell was
equipped with electrodes in a point-plane configuration, the 250~nm-radius
tip having been etched from a tungsten wire,
with an electrode gap of 3 to 4~mm. After purging and evacuating the
lines with a scroll pump, up to 100~bar helium of N6.0 research grade
purity was introduced via stainless steel pipes. A discharge was
ignited using a Spellman high 
voltage power supply using negative voltages between 3 and 10~kV connected to the tip
electrode, and currents between 0.1~and 10 $\mu$A depending on the
thermodynamic phase of the helium. Fluorescence emitted from the
discharge region was collected by an achromatic lens, which also served as a high
pressure window of the cell to maximise detection efficiency. 
The fluorescence light was collimated by two further lenses
and then guided, via two adjustable, metal-coated 
mirrors and an f~=~150~mm achromatic lens, onto the entrance slit of an
Andor Technology Shamrock SR303i Czerny Turner spectrometer equipped
with a Peltier-cooled (-65$^\circ$C) CCD camera (Andor iDus DV420,
CCD-12855). A 1200~mm$^{-1}$ grating blazed at 500~nm was employed,
centred around 640~nm, and
high resolution spectra were recorded (at a resolution of 0.2~nm).

\begin{figure}
\centering
\resizebox{0.99\columnwidth}{!}
{
\includegraphics{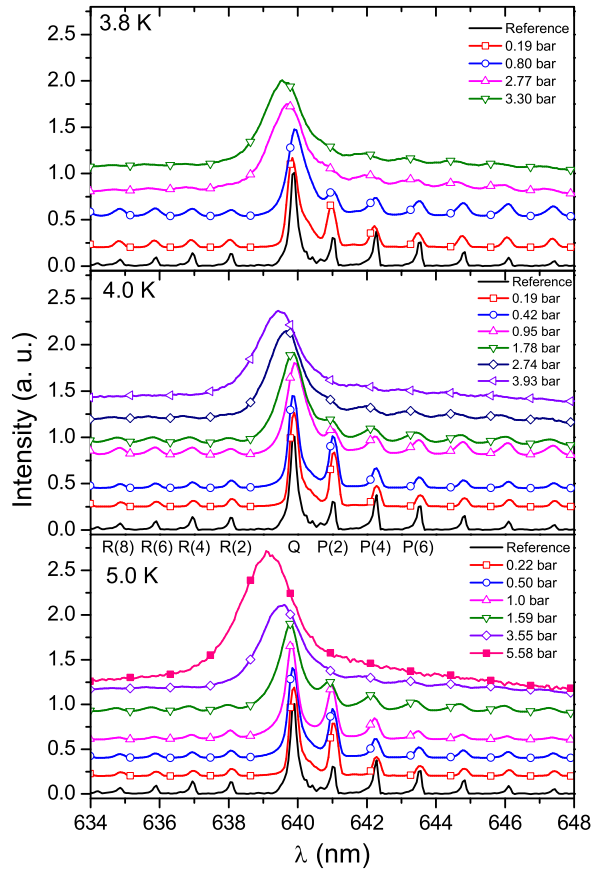}
}
\caption{
Spectra of He$_2^*$ in helium as a function of pressure for three
isotherms. The lines of a glow discharge reference spectrum are labelled R, Q, P as a function of quantum number $N$ (Hund's case (b)). With
increasing pressure, the lines broaden and shift. Note also that with increasing pressure
the Q and P(2) lines gain more intensity compared to other lines,
and ultimately merge. Intensities and offset have been scaled for better
visualisation. 
}\label{fig:spectra}
\end{figure}

Fluorescence spectra at 3.8, 4.0 and 5.0~K showing the
$d^3\Sigma_u^+ \rightarrow b^3\Pi_g$ transition of He$_2^*$ in the n~=~3
Rydberg state
\cite{ginter1965triplet} are shown
in figure~\ref{fig:spectra} for different hydrostatic pressures, covering
both the gas and the liquid phase of helium. A glow
discharge spectrum recorded in the evacuated cell is shown as reference
for both vacuum line positions \cite{ginter1965triplet} and the 
spectrometer's resolution. P and R lines are resolved; the Q lines lie
almost on top of each other and are thus too closely spaced to be resolved with our
spectrometer. All lines are split into groups of six energetically close
transitions due to spin-spin and spin-orbit interactions of the triplet levels in Hund's
case (b) \cite{fontana1962}. The triplet splitting also cannot be resolved at
our resolution.

At low cell pressures, sharp, discrete lines are observed. With
increasing pressure, the lines shift in frequency, broaden and change
their shape, particularly the P(2) and Q lines. By comparison with the
glow discharge spectrum, it can be clearly seen that the P(2) line
quickly gains intensity compared to the other P lines. Also, the P(2)
line merges at a certain, distinct pressure - for each given isotherm - with
the Q-branch. At 3.8~K this happens between 0.2 and 0.6~bar, at 4.0~K
between 0.7 and 0.9~bar, and at 5.0~K between 1.6 and 3.6~bar.
Rotational resolution vanishes when pressures increase much beyond
the SVP of {\em pure} helium, first for R-lines and then for P lines.

To analyse the changes in linewidth and line position, the lines were
separately fitted with Lorentzian functions. Over a large range of
pressures the lines remain symmetric, hence the fits with Lorentzians
represent a good way to assess and quantify line broadening by using the
spectral width as the sole parameter. To account for the
convolution of the widths of the Lorentzian lines with the spectrometer
response, 0.2~nm was subtracted from the fitted values, which is the
instrument resolution derived from the glow discharge spectrum.

\begin{figure}
\centering
\resizebox{0.99\columnwidth}{!}
{
\includegraphics{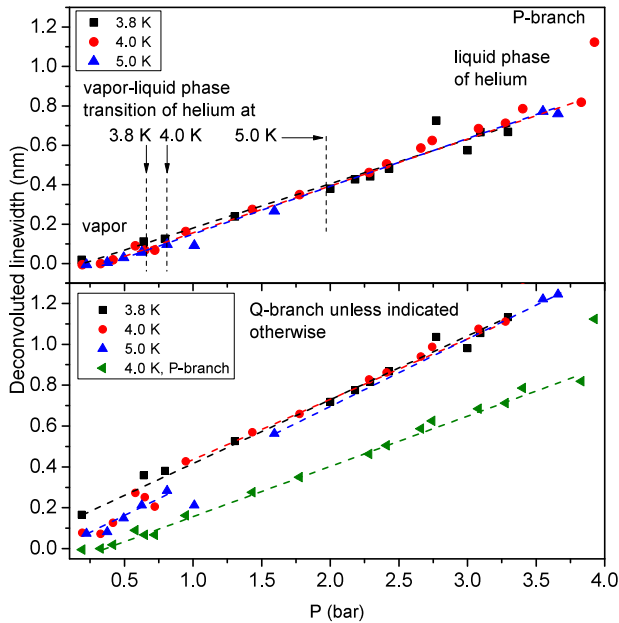}
}
\caption{
Pressure dependence of the linewidth of He$_2^*$ features (after
deconvolution, see text) for 3.8, 4.0 and 5.0~K isotherms. Upper panel:
analysis of P-lines. Lower panel: features in 
the region of the Q-transitions of He$_2^*$. The dashed lines show
trend lines. Results for P-transitions are shown for comparison. 
The line broadening of the 'Q-transitions' with pressure is
larger than for the P-lines.
}\label{fig:qp-widths}
\end{figure}

Figure~\ref{fig:qp-widths} shows the linewidths
obtained from the fitting procedure as a function of pressure; Q lines
are found to broaden more rapidly than P lines. 
At 3.8 and 4.0~K, the linewidth increases linearly with
pressure. Potential differences between the gas phase 
and the liquid phase of helium are too small to be resolved. 


\begin{figure}
\centering
\resizebox{0.99\columnwidth}{!}
{
\includegraphics{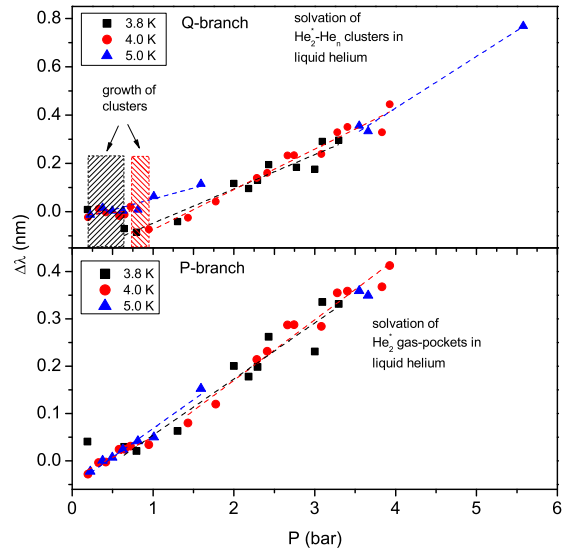}
}
\caption{
Line-shifts as obtained through fitting with Lorentzians. Upper panel:
`Q-transitions'. Lower panel: P-transitions. For the 3.8 and 4.0~K
isotherms the line-shifts differ greatly between Q and P transitions;
Q-features show larger line shifts with pressure than P-lines. Also,
abrupt changes (red shifts) indicating phase transitions are
observed. 
}\label{fig:q-lines}
\end{figure}

Figure~\ref{fig:q-lines} shows the Q- and P-lineshifts, $\Delta
\lambda$(p), as a function of
pressure. The lineshifts can be grouped into 
different pressure regions: 
at 3.8 and 4.0~K and pressures below the SVP of 
liquid helium, the Q-lines show small, and only slightly increasing shifts when the 
pressure is increased. 
The Q-lines then exhibit a small, but abrupt red shift. Further increases
of pressure then result in linearly, and  more greatly increasing, blue-shifts
than for lower pressures. The observed abrupt change indicates a 
change in the structure of the perturbing environment and is attributed
to the formation of a dense solvation shell. 
Lineshift coefficients, $\frac{\partial \Delta \lambda}{\partial p}$
and $\Delta \lambda_{0}$, were obtained for the gas and solvated phases
separately by fitting the function $\Delta
\lambda$(p) =  $\frac{\partial \Delta \lambda}{\partial p} p + \Delta
\lambda_{0}$ to the data and are shown in
table~\ref{tab:lineshiftconstants}, with the exception of the 3.8~K isotherm where
at pressures before the red shift, only one data point is available. At 4.0~K, the
gas-phase lineshift coefficient is 0.03~nm/bar, distinctly lower than the 0.17~nm/bar
found for the condensed phase.

Overall, the P-lines also show a linearly increasing blue shift. At 4.0~K, the
lineshift coefficient increases from 0.085~nm/bar to 0.128~nm/bar after
the SVP is crossed -- a change
that is smaller than that observed for the Q lines
(see table~\ref{tab:lineshiftconstants}).

\begin{table}[b]
\caption{\label{tab:lineshiftconstants}%
Line shift coefficients in the gas phase and liquid regions. 
}
\begin{ruledtabular}
\begin{tabular}{lllll}
  & & & \multicolumn{1}{c}{\textrm{slope, $\frac{\partial \Delta \lambda}{\partial p}$}} &
  \multicolumn{1}{c}{\textrm{intercept, $\Delta \lambda_{0}$}}\\
T & & & \multicolumn{1}{c}{\textrm{[nm/bar]}} & \multicolumn{1}{c}{\textrm{[nm]}}\\
\colrule
4.0~K & (Q) & gas & 0.03  $\pm$0.04 & -0.01  $\pm$0.03 \\
4.0~K & (P) & gas & 0.09  $\pm$0.01 & -0.03  $\pm$0.03 \\
5.0~K & (Q) & gas & 0.07  $\pm$0.02 & -0.03  $\pm$0.09 \\
\colrule
3.8~K & (Q) & liq. & 0.14  $\pm$0.01 & -0.19  $\pm$0.03 \\
4.0~K & (Q) & liq. & 0.17  $\pm$0.01 & -0.24  $\pm$0.03 \\
4.0~K & (P) & liq. & 0.13  $\pm$0.01 & -0.09  $\pm$0.03 \\
5.0~K & (Q) & liq. & 0.11  $\pm$0.02 & -0.07  $\pm$0.09 \\
\end{tabular}
\end{ruledtabular}
\end{table}

These observations indicate that the P and Q features must
originate from different species because in molecular spectra the
positions of P and Q lines are mutually dependent, as defined by the appropriate rotational constants (and hence, structure). 
We will see below that our
observations can be readily explained by a superposition
of spectra from excimers in two different types of environments:
He$_2^*$ residing (i) in a
solvated state and (ii) in hot gas pockets. Both species contribute, with
different spectral weights, to the Q and P lines of the spectrum. These
different contributions are apparent in the significantly larger
intensities of the Q and P(2) features compared to the P and R lines of
higher N in figure~\ref{fig:spectra}. For solvated, and hence cold,
He$_2^*$, only the lowest allowed quantum state, N~=~1, will be
populated because the next higher allowed level, N~=~3, is separated from it by
75~cm$^{-1}$. The only transitions emerging from the N~=~1 state are the
Q(1) and P(2) lines.


\begin{figure}
\centering
\resizebox{0.99\columnwidth}{!}
{
\includegraphics{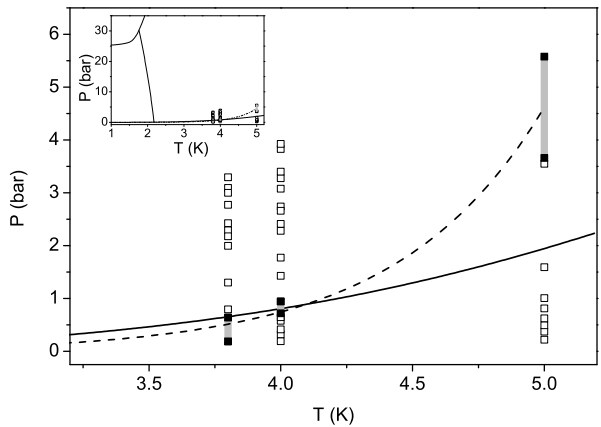}
}
\caption{
Phase diagram of normal liquid helium. 
The squares indicate measured data points on the 3.8, 4.0~K and 5.0~K
isotherms. The filled 
squares confine the region where the phase transitions of the
He$_2^*$-He system were observed; the dashed line is an exponential fit
to the mid-points.
Inset: overview section of helium phase
diagram with solid, liquid, superfluid, gas and supercritical phases. 
}\label{fig:phasediagram}
\end{figure}

\begin{figure}
\centering
\resizebox{0.99\columnwidth}{!}
{
\includegraphics{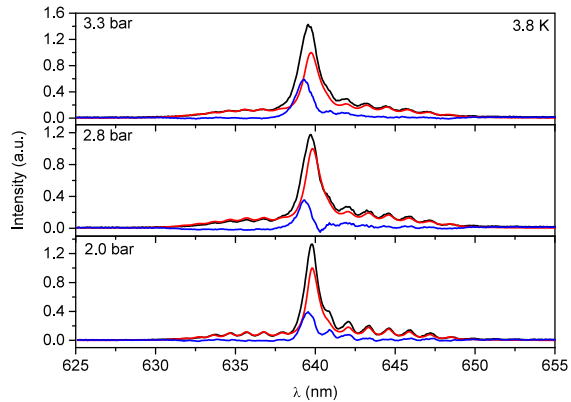}
}
\caption{
Illustration of how difference spectra were derived from a simulated
spectrum. The black line represents the measured spectrum; the red and
the blue lines represent the simulated and difference spectra,
respectively. The difference spectra show features in the region of the
Q-transitions that are offset to shorter wavelength, explaining how the
contribution from solvated excimers produces additional blue-shifting in
only this region.
}\label{fig:simulation}
\end{figure}

To disentangle the two contributions we have simulated the gas phase
spectrum and subtracted it from the measured features. The simulation
accounted for the triplet splitting of rotational levels, 
Boltzmann and H{\"o}nl-London factors 
and line broadening using a convolution of the lines
with a Lorentzian function. 
Figure~\ref{fig:simulation} illustrates this procedure for 2.0, 2.8 and
3.3~bar at 3.8~K.  
Best fits with the sharp line component were obtained for a rotational
temperature of 750~K and for higher hydrostatic pressures.
Significant intensity of the difference spectrum in the regions of the
P-lines shows that the assumed Boltzmann distribution is only
approximate and that the excimers are not strictly in equilibrium with a
thermal bath at 750~K. A temperature three orders of magnitude higher
than the solvent nevertheless indicates inefficient energy exchange with the
bulk helium environment, similar to excimers ejected from electronically
excited helium droplets \cite{haeften1997,haeften2002} and
excimers residing on the surface of helium droplets
\cite{yurgenson1999,mendoza2013}. The B-constants varied very little,
and always lay within 10\% the associated 
excimer gas phase values.

The difference spectrum revealed 20\% of the total intensity in
the region of the Q-transitions', with no rotational line structure,
suggesting that molecular rotation is strongly hindered, essentially as one would 
expect in normal liquid helium \cite{grebenev1998,sartakov2012}.
Supported by the resemblance to the spectrum of OCS in normal liquid $^3$He,
\cite{sartakov2012} we tentatively attribute the difference spectrum to the
envelope of a spectrum composed of the P(2) and Q(1) lines of He$_2^*$.

Summarising, our analysis of line broadening and lineshifts provides
evidence for He$_2^*$ residing in three different environments: (i) in
the gas phase, (ii) solvated and thermalised in clusters or
liquid helium, and (iii) in hot gas pockets within liquid helium. All
three states show different lineshift coefficients, the 'gas pocket'
coefficient lying between those of the gas and the condensed phases. The greater sensitivity to pressure changes means that the
perturbation of the He$_2^*$ in the gas pockets is larger than that in the
gas phase, indicating that the particle density inside the pockets is
rather high or, equivalently, the size of the pockets is rather small,
presumably only a little larger than the n~=~3 Rydberg-`bubble' diameter
of 15~\AA\ \cite{eloranta2001}. 

The observed discontinuities in the pressure-dependence of the lineshifts
can be readily explained as phase transitions of a mixed phase of
He$_2^*$ excimers and ground state helium atoms. 
We have marked the position of these discontinuities in the phase
diagram of helium in Figure~\ref{fig:phasediagram}. At 3.8~K this happens slightly below the SVP curve of {\em pure} helium, whilst at 4.0~K this happens {\em on} the {\em pure} helium SVP curve, whilst at 5.0~K this happens slightly above.
These results suggest an SVP curve for the He$_2^*$-He system which 
intersects the SVP curve of {\em pure} helium at 4.0~K.

This has the following implications: below 4.0~K, and in a region bounded by the SVP of {\em pure} and {\em
  mixed} helium, atoms nucleate on He$_2^*$ and form clusters. 
The fact that clusters of He$_2^*$ and He form before pure helium condenses
means that the He$_2^*$-He potential should have a shallow, long-range
minimum for the d~state. Such a minimum has not shown up yet in 
calculations \cite{eloranta2001}. 

Above 4.0~K, and between the SVP curves of {\em pure} and {\em
  mixed} helium, He$_2^*$-He mixtures can exist in gaseous form within
liquid helium. These gas bubbles are stable above the SVP of {\em pure}
helium within the liquid phase. This explains why sharp
lines were observed in the line spectra produced by corona discharge in
normal liquid helium \cite{li2009}. Furthermore, above 4.0~K, excess
energy from the corona discharge excitation leading to local heating can
be favourably released into He$_2^*$ residing in gas pockets, resulting
in He$_2^*$ excimers which are then able to boil within their own
solvation shell. 

He$_2^*$ reaching the surface of electronically excited helium
clusters desorb 
\cite{haeften1997,haeften2002,haeften2005ryd,bunermann2012} or remain bound to the surface
\cite{yurgenson1999,mendoza2013}. Rotationally resolved spectra show
that there are no helium atoms attached to the desorbed eximers
\cite{haeften1997,haeften2002}. Our findings suggest that desorption or
surface trapping depend on the temperature. Hence, investigation of the
respective transitions can provide insight into heating after
electronic excitation.   

In conclusion, we have investigated the spectra of He$_2^*$ in normal
liquid helium as a function of pressure and temperature. At low
pressure, rotationally resolved lines were observed. The lines shifted in
energy and broadened until they completely vanished as pressure was increased. Analysis of lineshifts and line broadening, and simulations of the He$_2^*$ gas phase
spectrum, show evidence for the presence of (i) cold excimers solvated
in helium, (ii) hot excimers in gas pockets within liquid helium and --
when the pressure is low enough -- (iii) in the gas phase. The excimers
rotate freely in the gas pockets and can thus display high rotational
temperatures of 750~K, while the solvated excimers are thermalised and
hindered in their rotation. Our work establishes a phase diagram for a
mixed phase of He$_2^*$ and ground-state helium, explaining the release of
energy by formation of microscopic gas bubbles, the desorption of
`naked' excimers from electronically excited helium clusters, and the
nucleation and stabilisation of He$_2^*$-He$_n$ at temperatures below
4~K.

\begin{acknowledgments}
KvH and FA acknowledge funding by the British Council
 through the Alliance Programme. KvH is grateful for financial
 support through The Leverhulme Trust (Research Grant F00212AH), the
 Royal Society (International Exchange Grant IE130173) and the
 Universit{\'e} Joseph Fourier for a Visiting Professorship.
LGML acknowledges financial support from the
Mexican Consejo Nacional de Ciencia y Tecnolog\'{i}a (CONACYT)
Scholarship number 310668, ID 215334. 
\end{acknowledgments}


\bibliography{abb_bibliography}
\end{document}